\documentclass[conference]{IEEEtran}
\usepackage{subfigure}

\ifCLASSINFOpdf
\else
\fi

\usepackage{algorithmic}
\usepackage{algorithm}
\usepackage{listings}

\usepackage{array}

\usepackage{stfloats}

\newcommand{\comment}[1]{}

\begin{document}
%
\title{Bounded Model Checking and Feature Omission Diversity}

\author{\IEEEauthorblockN{Mohammad Amin Alipour and Alex Groce}
\IEEEauthorblockA{School of Electrical Engineering and Computer Science\\
Oregon State University, Corvallis, OR\\
Email: \{alipour,alex\}@eecs.oregonstate.edu}
}


%


\maketitle

\lstset{ %
language=c,                
basicstyle=\footnotesize}

\begin{abstract}
In this paper we introduce a novel way to speed up the discovery of
counterexamples in bounded model checking, based on parallel runs over
versions of a system in which features have been randomly disabled. As
shown in previous work, adding constraints to a bounded model checking
problem can reduce the size of the verification problem and dramatically
decrease the time required to find counterexample. Adapting a
technique developed in software testing to this problem provides a
simple way to produce useful partial verification problems, with a
resulting decrease in average time until a counterexample is
produced. If no counterexample is found, partial verification results
can also be useful in practice.
\end{abstract}

\IEEEpeerreviewmaketitle

\section{Introduction}
\label{sec:intro}

Model checking \cite{ModelChecking} is a technique for formal
verification, which inspects all reachable states of a system for
violations of a property. If a violation found, a model checker
returns a counterexample---a trace that exhibits the property
violation.  Providing a counterexample is one of the most important
features of model checking, crucial for debugging both systems and
properties \cite{CountWitness}.


In programs with very large or infinite state spaces, the explicit
enumeration of all of reachable states is impractical.  Alternative
techniques, such as symbolic model checking \cite{McMillan1992}, model
checking with abstraction \cite{Clarke1994}, and \emph{bounded model
checking (BMC)} \cite{Biere1999}, avoid the need for explicit enumeration by
representing multiple states at once via logical constructs.  In
bounded model checking, in particular, there is an additional emphasis
on the discovery of counterexamples, and verification is only with
respect to some upper bound on the length of allowed counterexample traces\footnote{In practice, the bound can sometimes be shown to be at equal to or greater than that of the maximum length for the shortest counterexample to the system, in which case BMC verification implies system correctness.}.


Even with the power of symbolic techniques, the state-space explosion
is a fundamental problem for model checking: real-world systems have
very large state spaces, and effective abstractions and symbolic
representations are challenging to discover.  With the growing
prevalence of multi-core processors, effective methods for
parallelizing model checking have become an increasingly important
part of the effort to combat the state-space explosion. \emph{Swarm
verification} \cite{swarmIEEE} is an explicit-state approach in which,
rather than allocating all computational resources to a single model
checking search strategy (e.g., a bitstate search with a very large
hash table), a ``swarm'' of model checking runs with limited memory
and time, and different search strategies (primarily transition
orderings) are performed.  These runs are completely independent, with
no communication between processes; the cost of communication can
outweigh the gains of avoiding duplicated work.  The diversification
of search strategies leads to exploration of different parts of the
state space, however, decreasing the time required to find a
counterexample if one exists, and increasing code and state-space
coverage even if a counterexample is not found.  For very large
models, including an experimental network architecture, a real-time
operating system kernel, and flash file systems for space missions,
swarm verification has proven essential to finding subtle system
flaws.

 \emph{Swarm testing} \cite{Groce2011} generalizes the search
 diversification of swarm verification beyond the choice of a
 depth-first search strategy to the selection of \emph{test
 features}. A test feature is a predicate over test cases, controlled
 by the test generation process.  Features may be API calls or more
 general predicates; for example, in testing a file system a test
 feature might be whether the test case includes calls to {\tt close},
 and in C compiler testing a feature might be whether a test case
 includes pointer operations.  Building off the discovery that
 disabling certain API calls (e.g., directory operations) in an
 explicit-state model checking harness for a flash file system could
 greatly increase path coverage for other operations (e.g., file
 read/write operations), swarm testing relies on \emph{randomly
 omitting certain features from test cases}.  In the context of random
 testing, this means that, rather than using the entire test
 computation budget to generate and execute test cases drawn from a
 single distribution in which every test feature is represented, we
 generate a fixed swarm of test configurations, each of which only
 includes a subset of all possible test features.  The swarm
 configurations each receive an equal share of the computational
 budget.  The \emph{feature omission diversity} provided by swarm
 testing results in much better fault detection and code coverage than
 the conventional approach of a single, all-encompassing, default
 configuration.  Even though swarm testing, unlike swarm verification,
 provides no additional parallelism, diversity is valuable due to the
 fact that \emph{some features can suppress system behaviors}.  E.g.,
 when looking for a stack overflow bug, {\tt pop} operations make it
 hard to reach error states, and when testing a file system, {\tt
 close} operations prevent complex file access behaviors.

Swarm testing is clearly applicable in explicit-state model checking of
very large state spaces (in fact, the inspiration for swarm comes from
such ``testing via model checking'').  A more interesting question,
however, is whether feature omission diversity can be useful in
symbolic approaches.  In this paper, we propose the application of
swarm configuration techniques to constraint-based bounded model
checking.

Rather than simply using an \emph{ad hoc} approach of removing certain
choices from a model checking harness that defines system behaviors,
we base swarm bounded model checking (swarm BMC) on a trace-based
approach to bounded model checking \cite{Groce2008}. This approach
provides a generalized basis for feature omission in constraint-based
approaches, and swarm provides the trace-based technique with a source
of traces. We show very preliminary results on small systems,
indicating that 1) feature omission diversity can speed up the
discovery of counterexamples in BMC when multiple cores are available
and 2) it is possible to use the results from swarm runs to aid
verification, even if no counterexamples exist.

\section{Related Work}

This work builds on the ideas of search diversity explored in the work
of Holzmann et al. \cite{swarmIEEE} and Dwyer et al. \cite{DwyerSearch}, and
continued in swarm testing \cite{Groce2011}.  Feature diversity is
implemented by equating features to log elements in a program trace,
based on the approach to trace-based BMC reduction introduced by Groce
and Joshi \cite{Groce2008}.  Arguably, swarm's model checking via
feature omission, where model checking results indicate correctness
(or a counterexample) for a subset of program behaviors is a dual of
\emph{conditional model checking}, proposed by Beyer et
al. \cite{Condition}, where conventional model checking can return
partial results.


\section{Bounded Model Checking with Feature Omission}
\label{sec:fod}
Bounded model checking (BMC) reduces the problem of model checking to
a satisfiability problem \cite{Biere1999}. For a system $C$ and a
property $p$, BMC unrolls the control structure of the system for $k$
steps and derives a propositional formula $f$ such that $f$ is
satisfiable if and only if there is a violation of $p$ in computations
of length ($\leq$) $k$ of $C$. The solution to $f$ defines a
counterexample for the property $p$ over $C$.  A SAT solver (or SMT
solver) is used to solve the satisfiability of $f$.  If a sufficiently
large $k$ is selected and the SAT solver can handle the resulting
constraints, BMC is highly effective for finding faults in both
hardware and complex software systems, including operating system
kernel code.

Groce and Joshi \cite{Groce2008} proposed an extension to BMC in which
the only counterexamples considered are those that would produce a
given log tracing system behavior.  Consider a program containing {\tt
log(s)} statements that append a string {\tt s} to a log that
partially indicates program behavior.  Such logs are ubiquitous;
``{\tt printf}'' debugging is applied at some point to almost all
software systems. An obvious variation of bounded model checking is to
consider only counterexamples that produce a given log of system
behavior (perhaps hypothesized as impossible by a systems engineer, or
perhaps derived from a failed test case).  Restricting the executions
considered in a static analysis or symbolic execution by a trace can
be very expensive, if the restriction requires additional ``history''
variables to record the log produced by an execution.  In order to
avoid this overhead, the trace-restriction algorithm, after adding the
desired log state as a post-condition at program termination,
propagates the final assumption to arrive at a new program annotated
with assumptions that force the desired log output, and slices the
constraint problem according to these assumptions.  Experimental
results show that restricting behavior to a trace can greatly reduce
the size of SAT constraints and the time required to produce a
counterexample.

We exploit this approach as a basis for feature omission diversity. In
particular, we define features, in our swarm BMC approach, as logging
events.  To omit a feature, therefore, means to consider only
execution traces that do not contain a given log event.  This
definition of features is quite general: because logging statements
can be guarded with conditionals in a program, features can include
not only particular functions, statements, or input values, but also
any variable valuations at particular program locations.  Given a
program containing {\tt log} statements and a set of features to be
omitted (the swarm configuration), performing BMC under the
configuration is simple: 1) First, all logging statements that do not
involve the omitted features are removed from the program. 2) The
assumption that the program log is empty at termination is added as a
post-condition.  Restriction of execution to an empty log is most
elegantly achieved by simply replacing all calls to {\tt log} with the
statement {\tt assume(false)}.  These {\tt assume}s can be propagated
via the techniques described by Groce and Joshi, or by any slicing
technique implemented in a bounded model checker.

Algorithm \ref{alg:fod} illustrates the feature diversity omission
approach.  Given a program and a set of features, some features of
the program are randomly selected (Line \ref{alg:pick}).
These features are omitted from the program by adding new
assumptions to the program (Lines \ref{alg:omission} and
\ref{alg:slice}) to derive a set of new programs with the same behavior as the original program, except that none of the omitted features will be allowed in counterexamples.  We run BMC in parallel on the
set of new programs (Line \ref{alg:run}).

What is the value of this ``parallel'' BMC?  Obviously, there is no
traditional compositional verification in this case: if no $sp$ has a
counterexample, it does not follow that the program $p$ has no
counterexample (unless one $F_i$ is the null set).  However, if any
$sp$ has a counterexample, it is also a counterexample for the
original program $p$.  If the time to produce a counterexample for any
$sp$ is shorter than the time required to produce a counterexample for
$p$ then we have obtained a counterexample more quickly than what is
possible with the traditional approach.  Moreover, it may be the case
that BMC for reasonable $k$ on $p$ itself produces a SAT instance that
is too complex, and the SAT solver exhausts memory or time available
for verification.  It many such cases, at least one $sp$ will produce
a counterexample, due to its smaller (due to slicing) or at least more
constrained SAT problem.  In these cases, swarm BMC makes effective
BMC possible when it was previously not feasible.

\renewcommand{\algorithmicrequire}{\textbf{Input:}}
\begin{algorithm}

\caption{Algorithm For Swarm Bounded Model Checking}
\label{alg:fod}
\begin{algorithmic}[1]
\REQUIRE program $p$, set $F$ of features
\WHILE { budget allows }
	\FOR {all processors available} \label{alg:parallel}
			\STATE Pick a random set $F_i \subseteq F$ \label{alg:pick}
			\STATE Build $sp$ by replacing log statements in $p$ for $F_i$ with {\tt assume(false)} and removing other {\tt log} statements from $p$ \label{alg:omission}
			\STATE Propagate assumptions and slice $sp$ \label{alg:slice}
			\STATE Perform BMC on $sp$ \label{alg:run}
	\ENDFOR

\ENDWHILE
\end{algorithmic}
\end{algorithm}

We illustrate the potential value of the swarm approach with a simple
example, a stack overflow bug. Figure \ref{code:stack} depicts an
implementation of a simple stack API where {\tt push} adds an
integer to the top of stack, {\tt pop} removes an item from top of
stack and {\tt top} returns the value of the top of the stack. The
stack library is flawed in that 1) {\tt top} actually returns the value one
above the top of the stack and 2) the {\tt push} function does not check for
stack overflow.  The lack of a specification prevents us from
detecting the first fault, but the second fault will produce an array
bounds violation when a {\tt push} call is made on a stack already
containing 64 items.  The API calls are the natural features of this
system.

\begin{figure}
\begin{lstlisting}

#define SIZE 64
#define TLEN 100
int s = 0;
int stack[SIZE];
int top() {
  log("top");
  return stack[s];
}
void push(int i) {
  log("push");
  stack[s++] = i;
}
void pop() {
  log("pop");
  if (s > 0) {
    s--;
  }
}

int main () {
  int v, action;
  for (int i = 0; i < TLEN; i++) {
    action = nondet_int();
    __CPROVER_assume ((action >= 0) && (action <= 2));
    switch (action) {
    case 0:
      v = top();
      break;
    case 1:
      v = nondet_int();
      push(v);
      break;
    case 2:
      pop();
      break;
    }
  }
}
\end{lstlisting}
\caption{Stack code.}
\label{code:stack}
\end{figure}

The effectiveness of swarm testing and swarm BMC relies on the insight
that, whether features are API calls or predicates over inputs, most
faults can be exhibited by counterexamples that only exhibit a small
portion of program's behavior.  For example, in our stack example,
only {\tt push} operations are required to produce a stack overflow
--- {\tt pop} operations actually delay failure, and {\tt top} operations are irrelevant.

\if false
It took CBMC 325 seconds (without optimization) and 49 seconds (with optimization) to find a counter-example on a Pentium 4 2.8GHz. In this model, there are $3^{65}$ 65-long computation path exist that only one of them leads to the error state.
\fi

\begin{figure}[H]
\begin{lstlisting}
int top() {
  __CPROVER_assume(false);
  return stack[s];
}
void push(int i) {
  stack[s++] = i;
}
void pop() {
  if (s > 0) {
    s--;
  }
}
\end{lstlisting}
\caption{Snippet of input to CBMC to omit \textit{top} function calls.}
\label{code:cbmc1}
\end{figure}

\begin{figure*}
\centering
\begin{tabular}{|c | c| c | c |}
\hline
Omitted Feature & Time without Slicing (Seconds) & Time with Slicing (Seconds) & Verification Status\\
\hline
none & 325 & 49 & Counterexample\\
\hline
push & 40 & 4 & Verified\\
\hline
pop & 101 & 14 & Counterexample\\
\hline
top & 291 & 48 & Counterexample\\
\hline
\end{tabular}

\caption{Swarm BMC for a simple stack.}
\label{result:stack}
\end{figure*}

We ran the C bounded model checker CBMC \cite{CBMCp} on this program
and on three variations, each of which omitted one feature (Figure
\ref{code:cbmc1} shows the version where {\tt top} is omitted).
Figure \ref{result:stack} summarizes the results.

The results show that swarm BMC can returns a counterexample up to
71\% more quickly than BMC on the program with all features enabled.
Omitting {\tt top} provides little benefit, but omitting {\tt pop}
allows us to produce a counterexample very quickly.  If all BMC runs
are performed in parallel on a multi-core machine with sufficient
memory for each run, or in a cloud computing setting, the
time-to-first counterexample is greatly reduced.  Moreover, we argue
that counterexamples from swarm BMC are likely to be much more useful
for debugging systems, as they will contain fewer features that are
not relevant to the fault.  In this trivial case, a counterexample
consisting exclusively of {\tt push} operations is clearly ideal.

What if we correct the program by fixing {\tt top} and {\tt push}?
The swarm verifications are still (slightly) quicker than the full
verification, but do not provide a conclusive proof of correctness for
the program.  However, we can theorize some benefit even in the case
of correct systems for swarm verification: verifying the original
program takes three seconds longer than \emph{verifying a version of
$p$ which requires that every trace include at least one call to each
of {\tt top}, {\tt pop}, and {\tt push}}.

\section{Experimental Result}
\label{sec:experiments}

In this section we present preliminary results for swarm BMC. We have
experimented with feature omission diversity on a number of data
structures in C. In these experiments, we used only the basic CBMC slicer, rather than any more sophisticated reduction approach.

We used Weiss's \cite{Weiss1996} algorithms text source
\footnote{Available at {\tt http://users.cis.fiu.edu/$ \sim $weiss/
dsaa\_c2e/files.html}} as a source for simple examples. We modified
the source code for programs to introduce array bounds and null
pointer dereference faults.

We chose two data structures of this set: Array-Queue and Stack List.
The main program for each experiments invokes the functions of the libraries in a harness similar to the main function in Figure \ref{code:stack}.

We used CBMC version 4.0 \cite{CBMCp} for bounded model checking on a
four-processor Intel 2.8GHz system with 8 GB RAM.  Depth of bounded
model checking in the following experiments is the maximum number of
steps allowed in a trace.  In each experiment, we tried CBMC with and
without slicing, to show that simply adding additional constraints based on feature omission is valuable, even if no slicing is performed to reduce the SAT problem size.

\subsection{Array-Queue}

Array-Queue is an implementation of a queue in C using
arrays. Array-Queue includes {\tt Enqueue}, {\tt isEmpty},
{\tt Dequeue}, {\tt Front} and {\tt DisposeQueue} functions.
We introduced two bugs into the program: (1) violation of array
boundary caused by Enqueue-ing more data than the size of the queue,
and (2) null pointer dereferencing when calling an operation on a
disposed queue.

Figure \ref{res:arrayqueue} shows the results of bounded model
checking for Array-Queue. It shows the results for five different
depths: 1000, 2000, 3000, 4000, and 5000.  In all experimental
results, the fastest time to produce a failure is shown in bold.  In
every case, with or without slicing, all swarm configurations help
CBMC find a counterexample faster than with the default configuration.
Additionally, if the swarm configurations can be executed in parallel
(not unlikely; none of these runs required more than about 1GB of
RAM), swarm will produce two counterexamples showing different faults
before the default configuration can produce a single counterexample.

\begin{figure}
\centering

\subfigure[Without Slicing]{
\begin{tabular}{|c|c|c|}
\hline
Omitted feature & Depth & Time(S)  \\ \hline

\hline
\hline

 & 1000 & 22.979\\ \hline
Enqueue() & 1000 & 14.019\\ \hline
IsEmpty & 1000 & 17.930\\ \hline
{\bf Dequeue} & {\bf 1000} & {\bf 12.791}\\ \hline
Front & 1000 & 20.739\\ \hline

\hline
\hline

 & 2000 & 38.030\\ \hline
Enqueue() & 2000 & 24.125\\ \hline
IsEmpty & 2000 & 25.260\\ \hline
Dequeue & 2000 & 26.165\\ \hline
Front & 2000 & 24.425\\ \hline

\hline
\hline

 & 3000 & 38.067 \\ \hline
{\bf Enqueue()} & {\bf 3000} & {\bf 23.875} \\ \hline
IsEmpty & 3000 & 25.470 \\ \hline
Dequeue & 3000 & 26.245 \\ \hline
Front & 3000 & 24.316 \\ \hline

\hline
\hline

 & 4000 & 38.023\\ \hline
{\bf Enqueue()} & {\bf 4000} & {\bf 23.765}\\ \hline
IsEmpty & 4000 & 25.298\\ \hline
Dequeue & 4000 & 26.373\\ \hline
Front & 4000 & 24.282\\ \hline

\hline
\hline

 & 5000 & 36.966\\ \hline
{\bf Enqueue()} & {\bf 5000} & {\bf 23.382}\\ \hline
IsEmpty & 5000 & 24.918\\ \hline
Dequeue & 5000 & 26.193\\ \hline
Front & 5000 & 24.414\\ \hline

\hline
\hline

\end{tabular}

\label{arrayqueue:noslicing}
}

\subfigure[With Slicing]{
\begin{tabular}{|c|c|c|}
\hline
Omitted feature & Depth & Time(S)  \\ \hline
 & 1000 & 17.530\\ \hline
Enqueue() & 1000 & 13.352\\ \hline
{\bf IsEmpty} & {\bf 1000} & {\bf 12.639}\\ \hline
Dequeue & 1000 & 13.928\\ \hline
Front & 1000 & 13.048\\ \hline

\hline
\hline

 & 2000 & 	27.292\\ \hline
Enqueue() & 2000 & 	25.549\\ \hline
IsEmpty & 2000 & 	25.120\\ \hline
{\bf Dequeue} & {\bf 2000 }& {\bf 23.170}\\ \hline
Front & 2000 & 	27.826\\ \hline

\hline
\hline

 & 3000 & 	27.342\\ \hline
Enqueue() & 3000 & 	25.616\\ \hline
IsEmpty &  3000 & 25.526 \\ \hline
{\bf Dequeue} & {\bf 3000} & {\bf 22.594}\\ \hline
Front & 3000 & 	27.733\\ \hline

\hline
\hline

 & 4000 & 	27.196\\ \hline
Enqueue() & 4000 & 	25.821\\ \hline
IsEmpty & 4000 & 25.252 \\ \hline
{\bf Dequeue} & {\bf 4000} & {\bf 22.428}\\ \hline
Front & 4000 & 	27.433\\ \hline

\hline
\hline

 & 5000 & 	 26.615 \\ \hline
Enqueue() & 5000 & 	 24.591 \\ \hline
IsEmpty & 5000 & 	 24.735 \\ \hline
{\bf Dequeue} & {\bf 5000} & {\bf 21.678} \\ \hline
Front & 5000 &		 27.157\\ \hline

\hline
\hline

\end{tabular}
\label{arrayqueue:sliced}
}
\caption{Swarm BMC for Array Queue.}
\label{res:arrayqueue}
\end{figure}

\subsection{Stack List}
Stack List is a stack with dynamic memory allocation. It implements {\tt Push}, {\tt Top}, {\tt Pop}, and {\tt DisposeStack} functions. We added a null pointer dereferencing bug to the program.  Again, swarm BMC outperforms standard BMC, whether we choose a configuration at random and run on a single processor or we perform runs in parallel.

\begin{figure}
\centering

\subfigure[Without Slicing]{
\begin{tabular}{|c|c|c|}\hline
Omitted Feature & Depth & Time (S) \\ \hline

\hline
\hline

 & 100 & 0.291\\ \hline
Push & 100 & 0.283\\ \hline
{\bf Top} & {\bf 100} & {\bf 0.281}\\ \hline
Pop & 100 & 0.283\\ \hline
DisposeStack & 100 & 0.282\\ \hline

\hline
\hline

 & 200 & 1.592\\ \hline
{\bf Push} & {\bf 200} & {\bf 1.462}\\ \hline
Top & 200 & 1.498\\ \hline
Pop & 200 & 1.465\\ \hline
DisposeStack & 200 & 1.469\\ \hline

\hline
\hline

 & 300 & 7.151\\ \hline
{\bf Push} & {\bf 300} & {\bf 6.068}\\ \hline
Top & 300 & 6.472\\ \hline
Pop & 300 & 6.281\\ \hline
DisposeStack & 300 & 6.189\\ \hline

\hline
\hline

 & 400 & 28.509\\ \hline
Push & 400 & 25.646\\ \hline
Top & 400 & 26.468\\ \hline
{\bf Pop} & {\bf 400} & {\bf 19.392}\\ \hline
DisposeStack & 400 & 20.178\\ \hline

\hline
\hline

 & 500 & 69.376\\ \hline
Push & 500 & 70.351\\ \hline
{\bf Top} & {\bf 500} & {\bf 50.165}\\ \hline
Pop & 500 & 64.351\\ \hline
DisposeStack & 500 & 57.508\\ \hline

\end{tabular}
\label{stacklist:noslicing}
}

\subfigure[With Slicing]{

\begin{tabular}{|c|c|c|}\hline
Omitted Feature & Depth & Time (S) \\ \hline

\hline
\hline

 & 100 & 0.297\\ \hline
Push & 100 & 0.288\\ \hline
Top & 100 & 0.289\\ \hline
{\bf Pop} & {\bf 100} & {\bf 0.285}\\ \hline
DisposeStack & 100 & 0.287\\ \hline

\hline
\hline

 & 200 & 1.798\\ \hline
Push & 200 & 1.575\\ \hline
{\bf Top} & {\bf 200} & {\bf 1.593}\\ \hline
Pop & 200 & 1.595\\ \hline
DisposeStack & 200 & 1.659\\ \hline

\hline
\hline

 & 300 & 7.312\\ \hline
Push & 300 & 7.278\\ \hline
Top & 300 & 6.829\\ \hline
{\bf Pop} & {\bf 300} & {\bf 6.441}\\ \hline
DisposeStack & 300 & 6.825\\ \hline

\hline
\hline

 & 400 & 27.276\\ \hline
Push & 400 & 20.937\\ \hline
{\bf Top} & {\bf 400} & {\bf 20.092}\\ \hline
Pop & 400 & 21.188\\ \hline
DisposeStack & 400 & 20.484\\ \hline

\hline
\hline

 & 500 & 74.217\\ \hline
Push & 500 & 61.661\\ \hline
Top & 500 & 59.349\\ \hline
Pop & 500 & 60.722\\ \hline
{\bf DisposeStack} & {\bf 500} & {\bf 56.211}\\ \hline

\end{tabular}
\label{stacklist:sliced}
}
\caption{Swarm BMC for Stack List.}
\label{res:stacklist}

\end{figure}

\section{Discussion}
\label{sec:discussion}

In this paper, we proposed that feature omission diversity, known to be useful in software testing, may also be valuable in bounded model checking.
By omitting features in a program's execution, we can produce smaller and more easily checked SAT instances, while often preserving at least one counterexample trace.  This allows us to introduce parallelism into BMC without a parallel decision procedure for constraints.  The counterexamples produced, moreover, are potentially simpler to understand and debug than typical BMC counterexamples.

As future work, we plan to apply swarm BMC to larger, more realistic
examples that challenge the abilities of current software BMC.
Further investigation of the ability of specialized slicing to improve
runtime and the practicality of swarm parallel BMC are also needed.
Finally, we speculate that even when no counterexamples exist,
performing a number of swarm runs and then using the verified
configurations to restrict traces to consider in a run on the full
program may let us model check programs too large for verification
without additional constraints.

\bibliographystyle{acm}
\bibliography{cfv2011}
\end{document}